\documentclass[prd,superscriptaddress,showpacs,amsmath,%
preprintnumbers,showkeys]{revtex4-1}
\usepackage{graphicx}
\usepackage{epsf}
\usepackage{wrapfig}
\usepackage{epsfig}
\usepackage{epstopdf}
\newcommand{\beq}{\begin{equation}}
\newcommand{\eeq}{\end{equation}}
\newcommand\ba{\begin{eqnarray}}
\newcommand\be{\begin{equation}}
\newcommand\ee{\end{equation}}

\newcommand\ea{\end{eqnarray}}
\begin{document}
\title{Comments  on the observation of  high multiplicity events at the LHC}

\author{M~Strikman}
\address{Department of Physics,
 Pennsylvania
State University, University Park, PA 16802, USA
}
%\date{\today}
\begin{abstract}
We analyze the  structure of the high multiplicity events observed by the CMS collaboration at the LHC. We argue that the bulk of the observed correlations is due to the production of a pair of jets with $p_t > 15 $ GeV/c. 
We also suggest that high multiplicity events  are  due to a combination of three  effects: 
high underlying multiplicity for collisions at small impact parameters, upward fluctuations of the gluon density in the colliding protons, and production of hadrons  in the fragmentation of dijets.  The  data analysis is  suggested which may clarify the underlying  dynamics of the high multiplicity  events and probe  fluctuations of the gluon field as a function of $x$.
 \end{abstract}
\maketitle
\section{Introduction}
%\subsection{}
Recently the CMS collaboration  reported the first data  on the $pp$ collisions at $\sqrt{s}$= 7 TeV taken with a special  high multiplicity (HM)  trigger\cite{Khachatryan:2010gv}.  The trigger selected events with multiplicity $N_{ch}\ge 110$, in the pseudorapidity interval $|\eta| < $ 2.4 and $p_t > 0.4 \mbox{GeV/c}$. This corresponds to the multiplicities  $\sim$ 7 times larger than in the minimal bias inelastic events ($N_{min.bias}\sim 15$).
The reported studied the two-particle correlation  in pseudorapidity, $\Delta \eta=\eta_1-\eta_2$,  and  in the azimuthal angle $\Delta \phi=\phi_1-\phi_2$. 
The main focus of the discussion  was on the observation of the  same side positive correlation (ridge)
 at $\Delta \phi \sim 0$  in a wide rapidity interval of
   $2< |\Delta \eta | < 4$. A similar correlation was observed previously in the heavy ion collisions at RHIC.
 
 The CMS ridge effect was discussed in a number of theoretical papers. However little attention was payed to a number of  other remarkable features of the data. The aim of this note  is to fill this gap and to suggest  possible ways of  analyzing   the data which may clarify the dynamics of the HM events and provide a tool for studying fluctuations of the nucleon gluon density.

 Let us first summarize the basic observations of the CMS study.

\begin{itemize} 
\item The trigger selects events with 
average multiplicity which  is approximately seven times higher than for the minimum bias inelastic $pp$ collisions. Probability of the selected events is very  small,
\begin{equation}
P_{HM}\approx 10^{-5} \div 10^{-6}.
\label{hm}
\end{equation}
\item There is a very strong and  localized in $\Delta\eta$ 
positive two-particle correlation for $\Delta \phi \sim 0$ if  two particles with $1 < p_t < 3 $ GeV/c are selected.
 \item  For the same $p_t$ cut there is a  strong positive  two-particle correlation for $\Delta \phi \sim \pi $ for  a broad range of  $\Delta\eta$.
 
\item Total correlated hadron multiplicity  of hadrons in the same side   
$2 < |\Delta\eta | < 4.8$  ridge is $\sim $ 0.04 (0.02)
if  both selected  hadrons have comparable momenta in the range $1 < p_t < 2  (2 < p_t < 3) $  GeV/c  (the off-diagonal correlations were not reported yet). 
\end{itemize} 

\section{Origin of the away side ridge and  correlations in $\Delta \eta \sim 0, \Delta \phi \sim 0$}
The inspection of the correlation plot (Fig. 7 of ref. \cite{Khachatryan:2010gv})  indicates that the total excess transverse momentum in the $\Delta \phi \sim \pi$ region is  \begin{equation}
p_t^{balance} \ge \,   15 \mbox{GeV/c}.
\label{balance}
\end{equation}
It is not possible to estimate in the same way the  same side correlated transverse momentum since 
the  correlation function for $\Delta \phi =0 $ is off the vertical scale of the plot. However it is clear that the total excess  transverse momentum in the $\Delta \eta \sim 0, \Delta \phi \sim 0 $ region  is comparable to $p_t^{balance}$ (Eq.\ref{balance}).

Thus it appears  that the bulk of the observed 
$\Delta \eta, \Delta \phi $
correlations is due to production of two back to back jets with $p_t > $15 GeV/c.  This would explain both the  strong narrow correlation in the towards region and a broad correlation in the away region. The later is due to a broad distribution over  $x_1, x_2$ in the hard parton collisions at typical $x_i\sim 2p_t/\sqrt{s} \sim 5\cdot 10^{-3}$. A priori this does not exclude a possibility that  these jets are softer than the 
in-vacuum gluon  jets observed say in the $e^+e^-$ annihilation due to much higher  gluon densities characteristic for these collisions (see discussion below). This may be similar to the softening of the jets observed in the heavy ion collisions. Obviously, a softening of the jets  would make it rather difficult to extract  jets with such moderate transverse momenta from the data using existing jet finding  algorithms.

A natural question here is why $p_t^{balance}$ is so large. Naively the production of jets with smaller $p_t$  produced in the perturbative QCD (pQCD) regime with a higher  rate  should dominate
since the centrality for jets with different $p_t$ is practically the same \cite{Frankfurt:2003td,Frankfurt:2010ea} while the hadron multiplicity in the dijet fragmentation is a rather weak function of $p_t$.
 It is possible that a large value of $p_t^{balance}$  is related to the observation of the analysis \cite{Frankfurt:2010ea}  that $p_t$ at which pQCD starts to dominate grows with $\sqrt{s}$ and may be as large as 8 GeV/c in the generic  $\sqrt{s} $= 7 GeV collisions. Though no specific dynamic explanation of this pattern exists so far,  it appears likely that the pattern is  related to the increase of the gluon density in the $pp$ collisions with increase of  $\sqrt{s} $. If so, the effect should be stronger for collisions at small impact parameters  which as we argue below dominate the HM collisions since in this case the average gluon densities of the partons involved in the collisions are significantly higher than in generic inelastic collisions.
 
For the typical $\Delta \eta \sim 2$ we can estimate the invariant mass  of the produced dijet system
\begin{equation}
M_{jet_1\, jet_2} = 2 p_t  \exp(\Delta \eta/2 ) \sim 100 \mbox{GeV}.
\end{equation}

The multiplicities of hadrons in the fragmentation of the gluon jets were studied in a number of experimental papers, see 
 \cite{Abreu:1999rs} and references therein  - and found to be in  good agreement with the pQCD expectations \cite{book}.
   Using these data we can estimate that the charge hadron multiplicity  in these events due to the gluon dijet fragmentation is $\sim 30$, and hence gives a substantial ($\sim 25 \%$) contribution to the total multiplicity.
   
 Presence of  jets with large $p_t$ in the CMS  HM events may be of relevance for the interpretation of the same side large $|\Delta \eta|$ ridge. Indeed,  pQCD predicts a relatively small suppression of the  particle  density  (integrated over particle energies) emitted in-between jets. The suppression is   proportional to $\sqrt{N_c \alpha_s/\pi}$. Also, pQCD leads to the 
 difference between in- and out-of plane emissions  (a sort of "string effect"), in favor of in-plane emission  - a ridge like structure.  The actual numerical value is kinematics and color flow dependent (for extended discussion of the QCD coherence phenomena in the jet production see \cite{book}).
 
 It would be straightforward to find out  how  important  the production of  the back to back jets is in  the HM events with the data at hand.  This would allow also  to check 
  how much the pQCD "string effect" contributes to the same side ridge  (which constitutes $\sim 0.04 $ charged particle per event for $1< p_t < 2$ GeV/c). In particular one would be able to test predicted by pQCD dependence of the ridge on the the value of $\eta_1-\eta_2$.
     Comparison with the ridge structure in the minimal bias dijet  production for  $p_t \ge 20 GeV/c$ would be  useful  as well.
 
 \section{Dynamic mechanisms for generating high multiplicity
 and high $p_t$ jets}
 \label{sect:dynamic}

It is natural to expect that the average hadron multiplicity should monotonously increase with a decrease of the impact parameter, $b$. Hence the HM trigger of the CMS should correspond to the collisions  at very small $b$.
We can use the analysis of \cite{Frankfurt:2003td} to estimate the probability that an inelastic collision occurs  at small impact parameters. We find that the probability for   $b < 0.2$ fm  is  small $ \sim 2\%$ but still much larger than $P_{hm}$ (Eq.\ref{hm}).

Let us now estimate the average multiplicity for collisions at $b\sim 0$ for  $\sqrt{s}$ = 7 TeV.  The inclusive dijet trigger selects collisions 
at a median $b\approx$ 0.6 fm \cite{Frankfurt:2010ea}.
The  underlying multiplicity  for such collisions is about a factor of two larger than in the minimal bias non-diffractive events, see discussion and references in \cite{Frankfurt:2010ea,Aad:2010fh}. So for $b\sim 0$ the enhancement  should  be somewhat larger.

  A rough estimate  can be made using information about $b$-distribution of 
  minimal bias events for which median $b$ is $\sim 2$ than for the dijet events \cite{Frankfurt:2010ea}. 
  Experiments excluded diffractive events when calculating minimal bias hadron multiplicity. The  diffractive processes which constitute about 30\% of the total inelastic cross section  mostly occur at large impact parameters. 
  Taking this into account and comparing the $b$ distribution in  non-diffractive minimal bias events and in dijet events 
  one can "subtract" the large $b$ tail of the dijet distribution. As a result we estimate  that   the average multiplicity for  collisions  with $b < 0.6 fm$ is about  2.5 times higher than in minimal bias non-diffractive events. One may expect a further increase for $b\sim 0$.

 Hence we conclude that a scenario where small $P_{hm}$ is solely due to  the small probability of   small $b$ collisions and a small probability of selecting dijets with high $p_t$ would lead to a large multiplicity on the scale of  
 \begin{equation}
 N_{ch} = N_{jet jet} + N_{underlying}  \ge  70.
 \label{nch1}
 \end{equation}
  In this estimate  we have assumed for illustration that dijet multiplicity is of the order one.
   However the obtained $N_{ch}$   is still significantly smaller  than the observed one. Thus  HM trigger selects the tail of the distribution of the multiplicity  for central collisions.  
 
 It is natural to expect that to generate a higher multiplicity one needs to take into account  fluctuations of the strength of the gluon field on an event by event basis.   
 It was demonstrated in \cite{Frankfurt:2008vi} that one can relate the dispersion of the strength of the gluon field at small x to the ratio of inelastic and elastic vector meson production at $t=0$:
 \begin{equation}
\omega_g \;\; \equiv \;\; 
\frac{\langle G^2 \rangle - \langle G \rangle^2}{\langle G \rangle^2}
\;\; = \;\; 
\left[ \frac{d\sigma^{\gamma^\ast_L p 
\rightarrow VX}}{dt} \! \right/ \! \left.
\frac{d\sigma^{\gamma^\ast_L p 
\rightarrow Vp}}{dt} \right]_{t=0}.
\label{omega_g}
\end{equation}

A model of global fluctuations was proposed in \cite{Frankfurt:2008vi} which allowed the  explanation of  the magnitude of the experimental ratio in Eq. \ref{omega_g}. It  took into account the QCD DGLAP evolution of $\omega_g$.
For the discussed  CMS kinematics the model \cite{Frankfurt:2008vi} leads to  $\omega_g \sim 0.1$. 

We can define the probability for a fluctuation in a nucleon to have the strength of the gluon field  $\ge r$ times larger than  the gluon parton density which is the average over 
all configurations in the nucleon
as:
\begin{equation}
P(r)= \int d\sigma {G(x, Q^2 |\sigma)\over G(x, Q^2)} \theta(G(x,Q^2|\sigma) - r G(x,Q^2)),
\end{equation}
where $\sigma$ labels different configurations in nucleons.

Making a simplifying assumption that the distribution over the strength of the gluon field is Gaussian we can estimate 
\begin{equation}
P(1.5) \sim 3 \% .
\label{fluct}
\end{equation} 

However it is likely that the probability of  the large fluctuations is larger than given by  Eq.~\ref{fluct}. 
 Indeed the analysis \cite{Frankfurt:2003td} of the data on production of four jets ($\gamma$ + 3 jets) using  information about generalized parton distributions 
indicates presence of the positive multiparton correlations in nucleons.  Such correlations are likely  to increase $P(r\ge  1.5)$.

Hence an overall probability for the  interaction of two protons to occur at very small impact parameters $b < 0.2 $ fm and  with {\it both colliding nucleons}  in configurations having gluon density $ \ge 1.5$ times larger than   average, should be $\ge 10^{-5}$ which is comparable to the frequency of the events selected by  the HM trigger (Eq.~\ref{hm}). 
 This scenario would yield a larger value of $N_{ch}$ than the estimate of Eq.~\ref{nch1} due to a higher gluon density in the colliding configurations both due to soft and semi-hard interactions. Also it would increase the rate of binary parton collisions  per event.
 
 To estimate the rate of  dijet production in 2$\to$ 2 collisions we consider first the case of inclusive dijet production.  
 The normalized distribution over $b$ of the inclusive dijet production, $P_2(b)$   was calculated in \cite{Frankfurt:2003td,Frankfurt:2010ea} through the convolution of the  generalized gluon parton distributions which are measured in the exclusive hard processes at HERA. For the dipole parameterization of the two gluon form factor ($F_{2g}(t)  = (1 - t/m_g^2(x))^{-2}$):
 \begin{equation} 
  P_2(b)={m_g^2\over 12 \pi} \left( {m_gb\over 2}\right)^2 K_3(m_gb).
  \end{equation}
For the central collisions where $b\sim 0$ it is straightforward to find the ratio of the multiplicities in the $b\sim 0$  collisions and in the minimal bias collisions:
\begin{equation}
R_0= P_2(0)\sigma_{in}(pp)= {m_g^2\over 12 \pi}\sigma_{in}(pp).
\label{R0}
\end{equation}
For $x\sim 5\cdot 10^{-3}$, $M_g^2\approx 1 GeV^2$, leading to a significant enhancement of the dijet rate $R= 4.6$.

There are two types of fluctuations which modify $R$ - one is fluctuations of $g_N(x,Q^2|\sigma)$, and another is fluctuations of the size of the overlap area, $S$, of the colliding protons. For example, let us consider a model of a nucleon where the gluon field is mostly in the plane of three valence quarks - "a pancake shape". In this model $S$ for  the collisions when the  pancakes collide edge on edge is  much smaller than in average.
If we consider collisions in which  the three quarks in each of the nucleons are aligned along the reaction axis $S$ would be even smaller.
Since the rate of the jet production is $\propto 1/S$ such collisions  would lead to a large enhancement of the jet multiplicity.

Inclusion of the fluctuations would increase $R$ by a factor
$r_1\cdot r_2$:
\begin{equation}
R=R_0 r_1 r_2,
\end{equation}
  suggesting that the enhancement  could be as large as 10 for HM events, where
  \begin{equation}
  r_1r_2= {g_{N}(x_1,Q^2|\sigma)g_{1N}(x_2,Q^2|\sigma)\over g_{N}(x_1,Q^2)g_{1N}(x_2,Q^2)}
 { \left< S\right>\over S}
 \label{enhan}
 \end{equation}
  
  The multiparton 4$\to$ 4 hard collisions occur at even smaller impact parameters than  the 2$\to$ 2 hard collisions. Consequently, the 
   enhancement of the four  jet production  due to the   4$\to $ 4 interactions in HM events as compared to the minimal bias events should be larger than in the $2\to 2 $ case.   We can estimate the enhancement factor in the independent particle approximation using expression for $P_4(b)$ from \cite{Frankfurt:2003td}. Including the effect of fluctuations in the gluon density we find
  \begin{equation}
R_{4 jet}={7\over 35\pi} m_g^2 \sigma_{in} r_1r_2r_3r_4,
\end{equation}
which is several times larger than for the  two jet case.
   However the independent parton approximation underestimates the rate of four jet production\cite{Frankfurt:2003td},  so the enhancement factor may depend on the mechanism of parton-parton correlations and differ ffrom the square of the enhancement for the dijet production.
    
    \section{Testing color fluctuation conjecture}
    Above we have made a conjecture that the HM  
    trigger selects collisions of nucleons in configurations with larger than average gluon density. 
    In other words we suggest that HM and high gluon density
    fluctuations go hand in hand.
    
It is rather difficult to check this conjecture using the data analysis procedure of ref.  \cite{Khachatryan:2010gv} since one could always question whether the HM trigger  by itself was selecting events with extra hard collisions.

Here we discuss several possible strategies which address a possible dependence of the gluon fluctuations on $x$, virtuality of the gluon and overall transverse size of the projectiles.  Note here that so far the gluon fluctuations were modeled only in \cite{Frankfurt:2008vi}. This model  focused on the small $x$ region and led to a decrease of $\omega_{\sigma}$ with increase of $x$. In this model  significant 
 cancelations are present for the enhancement factor given by Eq. \ref{enhan} since in this model the upward fluctuation of the gluon field are related to the upward  fluctuations of the overall size  for configurations.
  However the model does not include fluctuations of the shape which may not change the gluon densities, for example "pancake" or "cigar" fluctuations discussed in the previous section. Such effects may lead to significant fluctuations  of $g_1g_2/S$  for a wide range of x.

One option to probe gluon field fluctuations  is to look for dijets at the central rapidities in matching back to back  narrow cones  but to use in the HM  trigger only hadrons outside these cones. This way one would be able to  determine how much  gluon fields in two nucleons are enhanced in the overlapping area as compared to the scenario where no fluctuations are present.

Another option is to consider the multiplicity of the dijet production, $N_{2j}$  in the region $|\eta| > 3.5 \div 3.0$  as a function of the multiplicity at central rapidities,$|\eta| < 2.4$ --  $N_{central}$.
The gap between the two regions  would minimize the  leaking of hadrons from hard collision to the central region. This problem is of importance only if $\eta_1 > 0, \eta_2<0$. Both the CMS and the ATLAS  
can measure jets in this kinematics using forward calorimeter components of their detectors.
In this kinematics one would test the fluctuations of gluons either with say  $x_1\sim 10^{-3}$ and $x_2 \gg x_1$ (same side jets) 
or  if $x_1\sim x_2 >  0.05$ (opposite side jets). 

One can expect that $N_{2j}$ ($N_{central}$) would grow monotonously 
with increase of $N_{central}$ due to increase of the centrality of the collisions. There should be a substantial increase of 
 $N_{2j}(N_{central} )$ between $N_{central}= N_{min. bias}$
 and $N_{central}= 2N_{min. bias}$ corresponding to typical 
 underlying multiplicity for dijet events (which reflects a more central nature of the  hard collisions). If the gluon  fluctuations would not kick in, $N_{2j}$ would saturate at the value 
 \begin{equation}
 N_{2j}= \lambda N_{2j} (N_{min.bias}),
 \end{equation}
 with $\lambda \sim  4$, cf. Eq.\ref{R0} with a small residual $x_i$ dependence due to the broadening of the transverse distribution of gluons with decrease of $x_i$ \cite{Frankfurt:2003td,Frankfurt:2010ea}
 
 Fluctuations may lead to a continued increase of 
 $N_{2j}$ for $N_{central} \gg 2 N_{min.bias}$ due to the fluctuations of the gluon field. The rate of the increase    may depend on $x_i$ and $p_t$ of the jets. In particular, in the model of  \cite{Frankfurt:2008vi} the increase of $N_{2j}$ for $b\sim 0$ events should be weaker for $\eta_1 \sim -\eta_2 \sim 3$ than for $\eta_1 \sim \eta_2 \sim 0$.
 
In principle, by combining measurements in three discussed kinematic ranges,  it may be possible to disentangle the x-dependence of the gluon fluctuations.

  \section{Conclusions}
 
 In the presented scenario HM events occur when the protons interact at very small impact parameter in special configurations with enhanced gluon field density. Production of two back to back jets with $p_t > 15 $ GeV/c in these events is strongly enhanced as compared to the minimal bias events.

   To test this hypothesis it is necessary to study the rate of   jet  production in the HM events in the vicinity of  the test  particle, and  the degree of correlation of such jets with a recoil jet production.
  Measurement of the ratio of the probability of high $p_t$ jets  in the HM at a wide range of rapidities  and minimal bias events would be especially revealing and may test the conjecture of the gluon fluctuations in nucleons and study x-dependence of the strength of these fluctuations.
  
  To check the relevance of the pQCD string  effect  for the explanation of  the same side ridge it is necessary to perform a comparison of the HM correlations with the correlations in  
  $pp$ scattering with production of dijets with $p_t\ge$ 20 GeV/c.  Alternatively the ridge could be due to a selection of rare configurations in the protons of a complicated transverse shape.
  
 \section*{Acknowledgements}
 I would like to thank J.~Bjorken, Yu.~Dokshitzer, L.~Frankfurt,  W.~Li, and C.~Weiss for many useful discussions. The work  has been  supported by DOE grant No. DE-FG02-93ER40771.

{\it Note edited in proof}. Very recently the ALICE collaboration reported preliminary data on the dependence of the $J/\psi$  multiplcity in pp collisions at $\sqrt{s}= 7 TeV$ on the hadron multiplicity at $\eta $ close to zero. For the highest measured hadron multiplicity which is about a factor of 5 larger than the minimal bias  multiplicity they report an increase of the $J/\psi$ multiplicity of about 
4  $\div $ 5. This is close to the estimate   of Eq.(9).

\end{document}